\documentclass[11pt]{article}
\usepackage{amsfonts,bm,bbm}
\usepackage{upgreek} 
\usepackage{mathrsfs}
\usepackage{amsmath}	
\usepackage{cite}
\usepackage{graphicx}
\usepackage{rotating}
\usepackage{hyperref}
\usepackage{verbatim}

\csname @addtoreset\endcsname{equation}{section}
\newcommand{\beq}{\begin{equation}}
\newcommand{\eeq}{\end{equation}}
\newcommand{\bea}{\begin{eqnarray}}
\newcommand{\eea}{\end{eqnarray}}
\newcommand{\ba}{\begin{array}}
\newcommand{\ea}{\end{array}}
\newcommand{\bit}{\begin{itemize}}
\newcommand{\eit}{\end{itemize}}
\newcommand{\nn}{\nonumber}

\newcommand{\mezzo}{\frac{1}{2}}
\newcommand{\complesso}{{\ \hbox{{\rm I}\kern-.6em\hbox{\bf C}}}}
\newcommand{\reale}{{\hbox{{\rm I}\kern-.2em\hbox{\rm R}}}}
\newcommand{\1}{ \,  \raisebox{+0.14em}{{\hbox{{\rm \scriptsize ]}} \raisebox{-0.2em}{\kern-.8em\hbox{1}}}} \, }  

\newcommand{\p}{\partial}

\renewcommand{\a}{\alpha}
\renewcommand{\b}{\beta}
\newcommand{\g}{\gamma}

\newcommand{\Er}{{\mathcal{E}}}

\renewcommand{\l}{\lambda}
\renewcommand{\L}{\Lambda}

\newcommand{\m}{\mu}

\newcommand{\n}{\nu}
\renewcommand{\r}{\rho}
\newcommand{\s}{\sigma}

\renewcommand{\t}{\theta}

\newcommand{\vf}{\varphi}

\newcommand{\om}{\omega}
\newcommand{\Om}{\Omega}

\begin{document}

\begin{titlepage}
\begin{flushright}
CECS-PHY-12/05
\end{flushright}
\vspace{2.5cm}
\begin{center}
\renewcommand{\thefootnote}{\fnsymbol{footnote}}
{\LARGE \bf Charging axisymmetric space-times}
\vskip 4mm
{\LARGE \bf with cosmological constant}
\vskip 20mm
{\large {Marco Astorino\footnote{marco.astorino@gmail.com}}}\\
\renewcommand{\thefootnote}{\arabic{footnote}}
\setcounter{footnote}{0}
\vskip 10mm
{\small \textit{
Centro de Estudios Cient\'{\i}ficos (CECs), Valdivia,\\ 
Chile\\}
}
\end{center}
\vspace{4.5 cm}
\begin{center}
{\bf Abstract}
\end{center}
{Ernst's solution generating technique for adding electromagnetic charge to axisymmetric space-times in general relativity is generalised in presence of the cosmological constant. Ernst equations for complex potentials are found and they are traced back to an affective dual complex dynamical system, whose symmetries are studied.\\
In particular this method is able to generate charged, asymptotically (A)dS black holes from their uncharged version: as an example, it is shown explicitly how to pass from the Kerr-(A)dS to the Kerr-Newman-(A)dS metric. A new solution describing a magnetic universe in presence of the cosmological constant is also generated.}
\end{titlepage}

\section{Introduction}

General Relativity (GR) with cosmological constant, nowadays, represents the standard model for the large scale interactions in our universe. Over the past years the cosmological constant ($\Lambda$) has been cheered both by a phenomenological and a theoretical point of view. It represents the actual best candidate for explaining dark energy and supernova accelerations, while theorists like because is an essential ingredient in the AdS/CFT correspondence (although they prefer a different positivity with respect the astronomers!). Exact solutions play a fundamental role in this scheme because they allow one to gain insight in the nature of gravity. For this reason exact solutions are ceaselessly studied and much effort has been devoted to their systematic construction. One of the most important classes of exact solutions in this contest is that of the axisymmetric ones. This class comprises static and stationary rotating black holes such as the ones found by Kottler (1918) \cite{kottler} and Carter (1968) \cite{carter_bh} known also as Schwarzschild-(A)dS and the Kerr-(A)dS metric respectively.\\
There is a wide literature about solution generating techniques in GR for asymptotically flat axisymmetric space-times. The most famous branches are the Ernst's complex potential method \cite{ernst1} - \cite{ernst2} (for a review see \cite{ernst3}) and the inverse scattering technique \cite{belinski}. \\
Ernst formalism revealed to be powerful enough to build several non-trivial solution such as black holes embedded in magnetic universe \cite{ernst-magnetic} both static, charged and rotating \cite{kerr-magnetic}. Moreover the Tomimatsu-Sato family of solution have been discovered and charged by this technique, as explained in\cite{ernst3}. These latters are known for $\Lambda=0$, and in this case they contain naked singularities, but when non-null cosmological constant is present, unexpected effects may occurs, such as the singularity may be hidden behind a horizon, so these metrics might describe black holes. This scenario is more likely in case of negative cosmological constant because uniqueness theorems are circumvented. Afterwards the generalisation of the Ernst method to five-dimension have granted even more involved solutions as multiple black rings \cite{di-ring}.  \\
Less is known about systematic construction in presence of cosmological constant: Some attempts to generalise \cite{ernst1} are proposed in \cite{charmo1}, but the problem of adding rotation to static axisymmetric space-times still remains open. In the present paper we address the issue of electro-magnetically charge axisymmetric space-times in the theory of four dimensional general relativity with the cosmological constant. In particular our aim is to generalise the Ernst's results of \cite{ernst2} to the case when the cosmological constant is present and to study the symmetry of the system. \\

In section \ref{complex-potencial} the general formalism for generic axisymmetric space-time is discussed, while in section \ref{ads-bh} a practical example is shown: how to add electromagnetic charge to the rotating (A)dS black hole, that is how to pass from the Kerr-(A)dS to the Kerr-Newman-(A)dS metric. Then in section \ref{magn-universe} a new class of solution is presented, they are interpreted as generalised magnetic universes in presence of the cosmological constant.

\section{Complex potential formalism}
\label{complex-potencial}

\subsection{Equations of motion}

Consider the action for general relativity with cosmological constant coupled to the Maxwell electromagnetic field:
\beq  \label{action}
                       I[g_{\m\n}, A_\m] =  \frac{1}{16 \pi}  \int d^4x  \sqrt{-g} \left[ \frac{1}{G} \left( R-2\L \right) - \frac{1}{\m_0} F_{\m\n} F^{\m\n} \right]  \ \ \ .
\eeq
Extremising the action with respect to the metric $g_{\m\n}$ yields the Einstein field equations, while extremising with respect to the electromagnetic potential $A_\m$ gives the Maxwell equations:
\bea  \label{field-eq}
                        &&   R_{\m\n} - \mezzo  R  g_{\m\n}  + \Lambda  g_{\m\n} = 2 \frac{G}{\m_0} \left( F_{\m\r}F_\n^{\ \r} - \frac{1}{4} g_{\m\n} F_{\r\s} F^{\r\s} \right)   \quad ,       \\
                        &&   \partial_\m ( \sqrt{-g} F^{\m\n}) = 0  \ \ \quad .
\eea
We now focus on stationary axisymmetric electromagnetic potentials $A=A_0dt+A_3d\varphi$ and space-times, whose metric can be generically written as:
\beq \label{axis-metric}
                          ds^2 = - \a e^{\Omega/2} \left( dt + \om d\varphi \right)^2 + \a e^{-\Omega/2} d\varphi^2 + \frac{e^{2 \nu}}{\sqrt{\alpha}} \left( d\text{r}^2 + d\text{z}^2 \right) \ ,
\eeq
where all functions $A_0$, $A_3$, $\Omega$, $\a$ , $\om$, and $\n$ depend just on (r,z) coordinates, which are not the standard spherical coordinates ($r,\vartheta$), but they are related to these latter by a change of coordinates. The operator $\overrightarrow{\mathbf{\nabla}} f=(\p_\text{r} f,\p_\text{z} f)$ has not to be confused with the usual gradient in three space dimensions. In terms of this ansatz the principal Einstein Equations (whose component will be labelled $EE_{\m\n}$) are:
\bea 
\hspace{-1cm}
     \label{eq-ee-rr-zz}         EE^{\text{r}}_{\ \text{r}} + EE^{\text{z}}_{\ \text{z}}  &:& \quad \nabla^2 \a + 2 \L \sqrt{\a} e^{2\n} = 0 \\  
     \label{eq-ee-p-t}                                  EE^{\vf}_{\  t}      &:& \quad \mezzo \overrightarrow{\nabla} \cdot \left( \text{e}^\Om \a \mathbf{\overrightarrow{\nabla}}\om \right) + 2\frac{G}{\m_0} e^{\Om/2} \left[ \om (\mathbf{\overrightarrow{\nabla}} A_0)^2 - \mathbf{\overrightarrow{\nabla}}A_0 \cdot \mathbf{\overrightarrow{\nabla}}A_3 \right] = 0 \\ 
             EE^{ t}_{\ t} - EE^{ \vf}_{\ \vf}  &:& \quad \mezzo \overrightarrow{\nabla} \cdot \left( \a \overrightarrow{\nabla}\Om \right) +  \om \overrightarrow{\nabla} \cdot \left( \text{e}^\Om \a \mathbf{\overrightarrow{\nabla}}\om \right) +\a e^\Om(\overrightarrow{\nabla}\om)^2+ \nn \\   \label{ee-eq-tt-pp}    & & \ \ + 2\frac{G}{\m_0} \left\{ e^{\Om/2} \left[ (\mathbf{\overrightarrow{\nabla}} A_0)^2 \om^2 - (\overrightarrow{\nabla} A_3)^2  \right] - e^{-\Om/2} (\overrightarrow{\nabla} A_0)^2 \right\} = 0 \\ 
            \label{eq-ee-tt+pp} EE^{ t}_{\ t} + EE^{ \vf}_{\ \vf}    &:& \quad \mezzo \a e^\Om (\overrightarrow{\nabla} \om)^2 - \frac{1}{8} \a (\overrightarrow{\nabla} \Om)^2  + \mezzo \nabla^2 \a  - 2\a \nabla^2 \n = 0 
\eea
While the Maxwell Equations ($ME$) read:
\bea
    \label{eq-em-t}  ME^t         &:& \quad    \overrightarrow{\nabla} \cdot \left[ e^{-\Om/2} \overrightarrow{\nabla} A_0 + \om e^{\Om/2} \left( \overrightarrow{\nabla} A_3 - \om \overrightarrow{\nabla} A_0 \right) \right] = 0    \\
    \label{eq-em-phi}     ME^\varphi   &:& \quad  \overrightarrow{\nabla} \cdot \left[ e^{\Om/2} \left( \overrightarrow{\nabla} A_3 - \om \overrightarrow{\nabla} A_0 \right) \right] = 0 \quad . 
\eea
Following the spirit of \cite{ernst1} and \cite{ernst2} equation (\ref{eq-em-phi}) may be regarded as the integrability condition for a magnetic scalar potential $\tilde{A}_3$ such that\footnote{The orthonormal frame is defined by the triad $(\overrightarrow{e}_\text{r},\overrightarrow{e}_\varphi,\overrightarrow{e}_\text{z})$}:
\beq \label{A3tilde}
      \overrightarrow{e}_\varphi \times \overrightarrow{\nabla}\tilde{A}_3 := e^{\Om/2} \left( \overrightarrow{\nabla} A_3 - \om \overrightarrow{\nabla} A_0 \right) \ \ . 
\eeq 
In fact from (\ref{eq-em-phi}) 
\beq
     \nabla \cdot \left( \overrightarrow{e}_\varphi \times \overrightarrow{\nabla} \tilde{A}_3 \right) = 0  \qquad \Longrightarrow \quad \p_\text{r} \p_\text{z} \tilde{A}_3 - \p_\text{z} \p_\text{r} \tilde{A}_3 = 0  \ .
\eeq
Moreover just using a vectorial identity from (\ref{A3tilde})
\beq
      \overrightarrow{e}_\varphi \times \overrightarrow{\nabla}A_3 = - \left( e^{-\Om/2} \overrightarrow{\nabla} \tilde{A}_3 - \om \overrightarrow{e}_\varphi \times \overrightarrow{\nabla} A_0  \right) \quad , 
\eeq
hence $ME^\varphi$ can be written in this form:
\beq \label{em-phi}
      \overrightarrow{\nabla} \cdot \left( e^{-\Om/2} \overrightarrow{\nabla} \tilde{A}_3 - \om \overrightarrow{e}_\varphi \times \overrightarrow{\nabla} A_0  \right) = 0 \quad .
\eeq
In view of the definition of $\tilde{A}_3$ in (\ref{A3tilde}) it is possible to rewrite $ME^t$ in (\ref{eq-em-t}) as
\beq  \label{em-t}
      \overrightarrow{\nabla} \cdot \left( e^{-\Om/2} \overrightarrow{\nabla}A_0 + \om \overrightarrow{e}_\varphi \times \overrightarrow{\nabla}\tilde{A}_3 \right) = 0 \quad .
\eeq 
Now one sees from (\ref{em-phi}) and (\ref{em-t}) that it is convenient to introduce a complex electric potential:
\beq \label{Phi-def}
      \mathbf{\Phi} := A_0 + i \ \tilde{A}_3 \qquad ,
\eeq
which, satisfying the equation
\beq \label{compl-em}
     \overrightarrow{\nabla} \cdot \left( e^{-\Om/2} \overrightarrow{\nabla} \mathbf{\Phi} - i \ \om \overrightarrow{e}_\varphi \times \overrightarrow{\nabla} \mathbf{\Phi} \right) = 0 \ \ ,
\eeq
allows us to cast all the Maxwell equations (\ref{eq-em-t})-(\ref{eq-em-phi}) in a unique complex equation (\ref{compl-em}). The real part correspond to the $ME^t$ while the imaginary part to the $ME^\varphi$.\\

Similarly we are going to treat the Einstein's equations\footnote{Where for convenience $G/\m_0$ is set to 1.}. For instance, thanks to the electric complex potential $\mathbf{\Phi}$, the $EE^\varphi_{\ t}$ (\ref{eq-ee-p-t}) can be written as:
\beq
     \overrightarrow{\nabla} \cdot \left[ e^{\Om} \a \overrightarrow{\nabla} \om - 2 \overrightarrow{e}_\varphi \times \text{Im} (\mathbf{\Phi^*} \overrightarrow{\nabla} \mathbf{\Phi} ) \right] = 0 \quad .
\eeq 
Again this equation can be considered as an integrability condition for the existence of a new potential $h$ such that
\beq
       \overrightarrow{e}_\varphi \times \overrightarrow{\nabla} h := e^{\Om} \a \overrightarrow{\nabla} \om - 2 \overrightarrow{e}_\varphi \times \text{Im} (\mathbf{\Phi^*} \overrightarrow{\nabla} \mathbf{\Phi} ) \quad .
\eeq
As for the electric case it is convenient to write the previous equation as
\beq \label{ephixdom}
      \overrightarrow{e}_\varphi \times \overrightarrow{\nabla} \om = - \a^{-1} e^{-\Om} \left[ \overrightarrow{\nabla} h + 2 \ \text{Im} (\mathbf{\Phi^*} \overrightarrow{\nabla} \mathbf{\Phi} ) \right] \qquad ,
\eeq
and hence we have the $EE^\varphi_{\ t}$ equation written in terms of $h$ and $\mathbf{\Phi}$ :
\beq \label{complex-ee-p-t}
 \overrightarrow{\nabla} \cdot \left\{ e^{-\Om} \a^{-1} \left[  \overrightarrow{\nabla} h + 2 \ \text{Im} (\mathbf{\Phi^*} \overrightarrow{\nabla} \mathbf{\Phi} ) \right] \right\} = 0 \quad .
\eeq
On the other hand the equation $EE^{ t}_{\ t} - EE^{ \vf}_{\ \vf}$ (\ref{ee-eq-tt-pp}) in terms $h$ and $\mathbf{\Phi}$ assumes the form
\beq \label{complex-ee-tt-pp}
     \frac{f}{\a} \overrightarrow{\nabla} \cdot \left( \a \overrightarrow{\nabla} f \right) - \overrightarrow{\nabla} f \cdot \overrightarrow{\nabla} f - f^2 \frac{\nabla^2 \a}{\a} = 2 f \overrightarrow{\nabla} \mathbf{\Phi} \cdot \overrightarrow{\nabla} \mathbf{\Phi^*} - \left[ \overrightarrow{\nabla} h + 2 \ \text{Im} (\mathbf{\Phi^*} \overrightarrow{\nabla} \mathbf{\Phi} ) \right]^2 \quad,
\eeq
where we defined $f:= e^{\Om/2} \a$. If one introduces the complex function
\beq \label{Er}
     \Er := f - | \mathbf{\Phi} |^2 + i \ h \qquad ,
\eeq
it is possible to write in a couple of complex equations both the Einstein equation $EE^{ \varphi}_{\ t}$ and $EE^{ t}_{\ t} - EE^{ \vf}_{\ \vf}$, and the Maxwell equations $ME^t$ and $ME^\varphi$:

\bea 
     \label{ee-ernst-lambda}  \left( \text{Re} \ \Er + | \mathbf{\Phi} |^2 \right) \frac{1}{\a} \overrightarrow{\nabla} \cdot \left( \a \overrightarrow{\nabla} \Er \right)    &=&   \left( \overrightarrow{\nabla} \Er + 2 \ \mathbf{\Phi^*} \overrightarrow{\nabla} \mathbf{\Phi} \right) \cdot \overrightarrow{\nabla} \Er + \text{Re}^2 \left( \Er + | \mathbf{\Phi} |^2 \right)  \frac{\nabla^2 \a}{\a}         \\
     \label{em-ernst-lambda}   \left( \text{Re} \ \Er + | \mathbf{\Phi} |^2 \right) \frac{1}{\a} \overrightarrow{\nabla} \cdot \left( \a \overrightarrow{\nabla} \mathbf{\Phi} \right)    &=& \left( \overrightarrow{\nabla} \Er + 2 \ \mathbf{\Phi^*} \overrightarrow{\nabla} \mathbf{\Phi} \right) \cdot \overrightarrow{\nabla} \mathbf{\Phi} \quad . 
\eea

To be more precise, the real and imaginary parts of (\ref{ee-ernst-lambda}) correspond to the (\ref{complex-ee-tt-pp}) and (\ref{complex-ee-p-t}) respectively; while the real and imaginary parts  of (\ref{em-ernst-lambda}) give $ME^t$ and $ME^\varphi$ respectively. \\
Equations (\ref{ee-ernst-lambda}) and (\ref{em-ernst-lambda}) represent the goal of this paper. Because they are the natural generalisation of Ernst's equations of \cite{ernst2} to the case with cosmological constant, so they reduce to the Ernst equations when $\Lambda \rightarrow 0$. In fact, by means of the equation (\ref{eq-ee-rr-zz}) the last term in (\ref{ee-ernst-lambda}) becomes null, and the function $\a$ becomes harmonic, $\nabla^2 \a(\text{r},\text{z}) = (\p^2_{\text{r}}+\p^2_{\text{z}})\a(\text{r},\text{z}) = 0 $, hence one can set $\a=\text{r}$. In the same way equations (\ref{ee-ernst-lambda}-\ref{em-ernst-lambda}) are the generalisation of the ones of \cite{charmo1} in presence of an electromagnetic field, and again they reduce to these ones when the electromagnetic field vanishes. \\

\subsection{Effective action and symmetries}

Remarkably enough equations for the complex potentials (\ref{ee-ernst-lambda}) - (\ref{em-ernst-lambda}) can be derived by an effective two dimensional action $S[\Er,\Er^*,\mathbf{\Phi},\mathbf{\Phi}^*]$:
\beq \hspace{-0.1cm} \label{maxw-GR+L-action}
     S = \int d\text{r} d\text{z} \ \a \left[ \frac{(\overrightarrow{\nabla}\Er + 2 \ \mathbf{\Phi^*} \overrightarrow{\nabla} \mathbf{\Phi} ) \cdot (\overrightarrow{\nabla}\Er^* + 2 \ \mathbf{\Phi} \overrightarrow{\nabla} \mathbf{\Phi}^* )}{(\Er + \Er^* + 2 \mathbf{\Phi} \mathbf{\Phi}^*)^2} - \frac{2 \overrightarrow{\nabla} \mathbf{\Phi} \cdot \overrightarrow{\nabla} \mathbf{\Phi}^* + \frac{\overrightarrow{\nabla}\a}{2\a} \cdot \overrightarrow{\nabla} \big( \Er + \Er^* + 2 \mathbf{\Phi} \mathbf{\Phi}^* \big)}{\Er + \Er^* + 2 \mathbf{\Phi} \mathbf{\Phi}^*} \right] 
\eeq

which, in the vacuum case, that is for $\mathbf{\Phi}=0$, reduces to:
\beq \label{action-GR+L}
     S = \int d\text{r} d\text{z} \ \a \left\{ \frac{\overrightarrow{\nabla}\Er \cdot \overrightarrow{\nabla}\Er^* }{(\Er + \Er^*)^2} -   \frac{\overrightarrow{\nabla}\a}{2\a} \cdot \frac{\overrightarrow{\nabla} \Er + \overrightarrow{\nabla}\Er^* }{\Er+\Er^*}  \right\} 
\eeq
Let's focus on this easier example first: When the cosmological constant is null the second term in (\ref{action-GR+L}) is absent. In that case we are left with pure GR, so it is well known that in this case the system enjoys a SL(2,$\mathbb{R}$) (or SU(1,1)) symmetry represented by generalised Ehlers transformations:
\beq
\Er \longrightarrow \Er' = \frac{a \Er + i b}{i c \Er + d} \ \ , \qquad \text{where} \qquad
\begin{pmatrix}
  a & b  \\
  c & d
\end{pmatrix}
\in SL(2,\mathbb{R}) \quad ,
\eeq
which explicitly look like
\bea
      i)    && \Er \longrightarrow \Er' = \l^2 \Er  \qquad \  \quad \ , \qquad \quad (a=\l ,\ b=0 ,\ c=0 ,\ d=1/\l)  \\
      ii)   && \Er \longrightarrow \Er' = \Er + i \ b \qquad \ , \qquad \quad (a=1,\ b=b, \ c=0 , \ d=1)   \\
      iii)  && \Er \longrightarrow \Er' = \frac{\Er}{1+ic\Er}   \qquad  , \qquad \quad (a=1,\ b=0,\ c=c, \ d=1)
\eea
When the cosmological constant is present $(i)-(ii)$ are still trivially respected, while $(iii)$ might hold only if $\a$ transforms (at least for some particular subclass of space-times) such that: 
\beq \label{trasf-alpha}
(\a -\a') \frac{\overrightarrow{\nabla}\Er \cdot \overrightarrow{\nabla}\Er^* }{(\Er + \Er^*)^2} = \mezzo (\overrightarrow{\nabla}\alpha - \overrightarrow{\nabla}\alpha') \frac{\overrightarrow{\nabla}\Er + \overrightarrow{\nabla}\Er^* }{\Er + \Er^*} - \overrightarrow{\nabla}\alpha' \frac{ic}{2} \left[ \frac{\overrightarrow{\nabla}\Er^* }{1-ic\Er^*} - \frac{\overrightarrow{\nabla}\Er^*}{1+ic\Er} \right]
\eeq
The change of the group of symmetry, in presence of the cosmological constant, may be the cause of failure of the naive application of Ernst's techniques for adding rotation to the static solution. Actually the symmetries of the equations of motion can be deduced by the homothetic symmetries (i.e.  the conformal Killing vectors, with constant conformal factor) of the three-dimensional metric associated to the action (\ref{action-GR+L}):
\beq
         ds^2 = \alpha \frac{dx^2-dy^2}{4x^2}- \frac{d\a \ dx}{2x} \qquad ,
\eeq
where the real coordinates of the metric are defined as follows: ($x$:=Re($\Er$) , $y$:=$i$Im($\Er$)).\\
Even though the independent homotetic Killing vectors remains three:
$$  \chi = \l x  \p_x + (b + \l y) \p_y + \tilde{c}\a\p_\a \ ,$$
 it is easy to see that they can generate only the symmetry trasformation $(i) , \ (ii)$ and a dilatation (of a factor $\tilde{c}$) in $\a$, while keeping $\Er$ unchanged. Hence, in this setting, the symmetry $(iii)$ is confirmed to be broken.

Similarly when the electromagnetic field is switched on and the cosmological constant is null, the group of symmetry is extended to SU(2,1), represented by the finite Kinnersley transformations:
\bea
      I)    && \Er \longrightarrow \Er' = \l \l^* \Er  \qquad \  \ \ \qquad \quad,  \qquad \mathbf{\Phi} \longrightarrow  \mathbf{\Phi}' = \l \mathbf{\Phi}   \\
      II)   && \Er \longrightarrow \Er' = \Er + i \ b \qquad \ \qquad \quad, \qquad \mathbf{\Phi} \longrightarrow  \mathbf{\Phi}' = \mathbf{\Phi}   \\
      III)  && \Er \longrightarrow \Er' = \Er/(1+ic\Er)   \qquad  \quad \ , \qquad \mathbf{\Phi} \longrightarrow  \mathbf{\Phi}' = \mathbf{\Phi}/(1+ic\Er)  \\
      IV)   && \Er \longrightarrow \Er' = \Er - 2\b^*\mathbf{\Phi} - \b\b^* \ \quad , \qquad \mathbf{\Phi} \longrightarrow  \mathbf{\Phi}' = \mathbf{\Phi} + \b   \\
      V)    && \Er \longrightarrow \Er' = \frac{\Er}{1-2\g^*\mathbf{\Phi}-\g\g^*\Er}  \ \  , \qquad \mathbf{\Phi} \longrightarrow  \mathbf{\Phi}' = \frac{\mathbf{\Phi}+\g\Er}{1-2\g^*\mathbf{\Phi}-\g\g^*\Er}\ , 
\eea
where $b, c \in \mathbb{R}$ and $\l,\b, \g \in \mathbb{C}$. When the cosmological constant is present $(I),(II),(IV)$ are trivially satisfied, but $(III),(V)$ are broken. In fact analysing the homothetic Killing symmetries of the five-dimensional metric associated to the action (\ref{maxw-GR+L-action}):
$$ds^2 = \a \frac{\frac{dx^2}{4}-\frac{dy^2}{4}+v(dvdx-dwdy)+ w(dvdy-dwdx) - x(dv^2-dw^2)}{(x+v^2-w^2)^2} - d\a \frac{dx + 2 v dv-2w dw}{2(x+v^2-w^2)}$$
it is possible to confirm that the only symmetries left are the ones of $(I),(II),(IV)$ and a trivial dilatation for $\alpha$. The coordinate $v$ and $w$ are defined such that $v$:=Re($\mathbf{\Phi}$) , $w$:=$i$Im($\mathbf{\Phi}$). \\
Since the action (\ref{maxw-GR+L-action}) does not enclose the whole set of equations of motion (\ref{eq-ee-rr-zz}) - (\ref{eq-em-phi}) one has to check that the possible group of homothetic symmetries of (\ref{maxw-GR+L-action}) are respected by eq. (\ref{eq-ee-rr-zz}) and (\ref{eq-ee-tt+pp}) as well. 

\section{Example: Electrifying the Kerr-(A)dS Black Hole}
\label{ads-bh}

In the previous section the Ernst complex potentials technique, casting the Einstein and Maxwell equations to a couple of complex equations, was generalised to the presence of the cosmological constant. In this section we will see, as an example, how to apply that insight to add electromagnetic charge to an axisymmetric exact solution of General Relativity with cosmological constant to get a vacuum of the same theory coupled to the Maxwell electromagnetism. In spite of the fact that some symmetries of the system have been broken by the presence of $\Lambda$ we cannot use a sort of generalised Kinnersley transformations, nevertheless the formalism developed so far will reveal somehow useful. This is done in the same spirit of Ernst \cite{ernst2}, who was able to generate the Kerr-Newman solution from the Kerr one without the help of the subsequent Kinnersley trasformations.

\subsection{Kerr-(A)dS metric}

 In particular we focus, as a starting point, on the metric found by Carter in \cite{carter_bh}, better known under the name of Kerr-(A)dS, describing a stationary, uncharged, rotating black hole with an asymptotically (Anti) de Sitter behaviour. This metric will represent our seed. In spherical coordinates ($r,\t,\varphi$) it reads\footnote{In order to have a well-behaved metric on the axis $\theta=0,\pi$, usually  $\varphi$ (and eventually $t$) is rescaled by an additional constant factor $\Xi=1+\L a^2/3$, which for the scope of this work is unnecessary, thus it is omitted.}:
\beq \label{kerr-ads}
     ds^2 = - \frac{\Delta_r}{\varrho^2} \left(dt - a \ \sin^2\theta \ d\varphi \right)^2 + \varrho^2 \left( \frac{dr^2}{\Delta_r} + \frac{d\t^2}{\Delta_\t} \right) + \frac{\Delta_\t \ \sin^2\t}{\varrho^2} \left[ a dt - (r^2+a^2)  d\varphi \right]^2
\eeq
where
\bea
     \varrho^2 &=& r^2 + a^2 \cos^2 \t  \nn \\
     \Delta_\t &=& 1 + \frac{\Lambda}{3} a^2 \cos^2 \t \nn \\
     \Delta_r &=& (r^2+a^2) \left(1-\frac{\Lambda}{3} r^2 \right) - 2Mr \quad . \nn 
\eea
First of all we need the Ernst complex potential for this metric. To find it one just needs, up to the definition (\ref{Er}), to determine the functions $\alpha,e^\Om,\om$, by comparing the generic axisymmetric metric (\ref{axis-metric}) with the Kerr-(A)dS one (\ref{kerr-ads}):
\bea
        \a &=&  \sin \t \sqrt{\Delta_r \Delta_\t}  \\ 
        e^{\Om/2} &=& \frac{\Delta_r - a^2 \Delta_\t \sin^2 \t}{\sqrt{\Delta_r \Delta_\t} \varrho^2 \sin \t}  \\
        \om &=&  \frac{a \sin^2 \t \left[ \Delta_r - \Delta_\t (r^2 + a^2) \right]}{a^2 \Delta_\t \sin^2 \t - \Delta_r}  \quad .
\eea
Then integrate (\ref{ephixdom}) to get $h$, taking into account that in this case the electromagnetic field is absent, so setting $\mathbf{\Phi}=0$. Thus it can be easily verified that the Ernst complex potential for the Kerr-(A)dS is
\beq  \label{ernst-pot-carter}
                                  \Er = \a e^{\Om/2} + i h = \frac{\xi - 1}{\xi + 1} + \frac{1}{\ell^2} \left[ (\xi+1)^2 + q^2 \right] \quad ,
\eeq
where we defined $1/\ell^2=\pm \L M^2/3$ (positivity depending on the sign of the cosmological constant $\L$) and we took as in \cite{ernst1}
$$ \xi = p x - i q y \quad, \quad p = \frac{k}{M} \quad , \quad x = \frac{r-M}{k} \quad , \quad q=\frac{a}{M} \quad , \quad y = \cos \t \quad, \quad k=\sqrt{M^2-a^2} \quad .$$
Actually can it be shown, by substitution, that this potential is a solution of the Einstein equations (\ref{ee-ernst-lambda}) in the case of vanishing electric and magnetic charges, that is when $\mathbf{\Phi} \rightarrow 0$ (of course in that limit the (\ref{em-ernst-lambda}) equations are trivially null). $\Er$ is solution whenever the real constants $p,q$ satisfy the relation $p^2+q^2=1$, as in the case without cosmological constant \cite{ernst1} - \cite{ernst3}. Note that for the static case, the Schwarzschild-(A)dS black hole, it is sufficient to switch off the rotational parameter, fixing $q=0$. However the presence of the cosmological constant, at least in this set of prolate spherical coordinates, somehow breaks the symmetry between the static and the stationary metrics, as the term $q^2/\ell^2$ in (\ref{ernst-pot-carter}) suggests. In \cite{charmo1} is claimed that this symmetry could be restored by a different set of coordinates. Anyway this means that is not possible to use this formalism, as it is, to add rotation to the Schwarzschild black hole in presence of the cosmological constant. But still our technique is useful to add electromagnetic charge to it.   \\
It is interesting to observe that the complex potential $\Er$ exactly reduces to the one of Ernst \cite{ernst1} in the limit of vanishing cosmological constant (or equivalently as $\ell \rightarrow \infty$). \\

\subsection{Charging Kerr-(A)dS metric}

Following the charging procedure of \cite{ernst2} and \cite{ernst3}, it is possible to find a solution for the general charged case, that is the complex system of equations (\ref{ee-ernst-lambda})-(\ref{em-ernst-lambda}). Primarily we need to stem the form of $\mathbf{\Phi}$; to do that one usually analyses the asymptotic behaviour and one assumes that $\Er$ is an analytic function of $\mathbf{\Phi}$.
In this way, by chain rule, it is possible to write  
$$  \frac{d^2 \Er}{d\mathbf{\Phi}^2} \ (\overrightarrow{\nabla} \mathbf{\Phi})^2 \ \overrightarrow{\nabla} \mathbf{\Phi} =  \nabla^2 \Er \ \overrightarrow{\nabla} \mathbf{\Phi} - \nabla^2 \mathbf{\Phi} \ \overrightarrow{\nabla} \Er \quad .$$
Then, using the equations (\ref{ee-ernst-lambda})-(\ref{em-ernst-lambda}), it follows that:
\beq \label{asympthotic-eq}
       \frac{d^2 \Er}{d\mathbf{\Phi}^2} ( \overrightarrow{\nabla} \mathbf{\Phi} )^2 = \left( \text{Re} \Er + |\mathbf{\Phi}|^2 \right) \frac{\nabla^2\a}{\a}  \quad .
\eeq
Now let us decompose the complex Ernst potential $\Er$ in terms of the cosmological constant: $\Er=\Er_0+\Er_\ell$, where $\Er_0$ is the zeroth order in the cosmological constant (does not contain it), while $\Er_\ell$ depends on $\ell$. Since on the right hand side of (\ref{asympthotic-eq}) there are no terms at the zero order in the cosmological constant we have: 
$$  \frac{d^2 \Er_0}{d\mathbf{\Phi}^2} = 0 \qquad \Longrightarrow \qquad \Er_0 (\mathbf{\Phi}) = c_0 + c_1 \ \mathbf{\Phi}  $$
The value of $\Er_0$ can be directly read from (\ref{ernst-pot-carter}): $\Er_0=(\xi-1)/(\xi+1)$.  
The arbitrary constant $c_1$ will be defined $c_1:=-2/Q$ for convenience, while $c_0$ can be fixed to one by the boundary condition at infinity: $\Er_0 \rightarrow 1$ and $\mathbf{\Phi} \rightarrow 0 $. Therefore is possible to derive the shape of the electromagnetic complex potential: 
\beq \label{Phi}
       \mathbf{\Phi} = \frac{Q}{\xi +1} \quad ,
\eeq
which naturally corresponds to the case when cosmological constant is null.\\
Thus, thanks to the (\ref{Er}) we are able to infer the value of $f$ and $h$ in the general case:
\bea
     \label{fQ}  f & = & \text{Re} (\Er + |\mathbf{\Phi}|^2)  =  \frac{\xi^* \xi - 1 + |Q|^2}{|\xi+1|^2} + \frac{1}{\ell^2} \ \text{Re} \left[ (\xi  + 1)^2  + q^2 \right] = \\
         & = &  1 - \frac{2 M r -e^2 -g^2}{r^2 + a^2 \cos^2 \t} + \frac{1}{\ell^2 M^2} \left[ r^2 + a^2 \sin^2 \t \right] = \frac{\Delta_e - a^2 \sin^2 \t \Delta_\t}{\varrho^2} \quad , \nn \\
       h & = & \text{Im} (\Er)  = \frac{2 \ \text{Im}(\xi)}{|\xi+1|^2} + \frac{1}{\ell^2} \ \text{Im} \left( \xi^2 + 2 \xi \right)  = - 2 a \cos \t \left( \frac{M}{r^2+a^2}+\frac{r}{\ell^2M^2} \right) \quad .
\eea
In eq (\ref{fQ}) we borrowed the standard definition of $Q:=(e+ig)/M$, where $e$ and $g$ are the electric and magnetic charges.
Observe that by definition of $\Er$ the imaginary part is not affected by the electromagnetic field, so that $h$ remains the same as that of the uncharged case. On the other hand the $f$ function remains of the same form of the uncharged case but the effect of the electromagnetic potential cause a shift of $\Delta_r$  to $ \Delta_e= \Delta_r + e^2 + g^2 $, while $\Delta_\t$ remains unchanged.     \\
Now we can plug $\mathbf{\Phi}$ and $h$ into the vectorial eq. (\ref{ephixdom}), which gives a first order, quasi-linear, homogeneous partial differential equation for $\om$ and an algebraic equations for $\a e^{\Om}$:
 \begin{equation*}
\left\{
\begin{array}{rl}
 &     \p_\text{z} \om \left[ \p_\text{z} h + 2 \ \text{Im}(\mathbf{\Phi^*} \p_\text{z} \mathbf{\Phi})) \right] = - \p_\text{r} \om \left[ \p_\text{r} h + 2 \ \text{Im}(\mathbf{\Phi^*} \p_\text{r} \mathbf{\Phi})) \right] \quad ,\\
 &   \a e^{\Om} = \left[ \p_\text{r} h + 2 \ \text{Im} (\mathbf{\Phi^*} \p_\text{r} \mathbf{\Phi}) \right] / \p_\text{z}\om \\
\end{array} \right.
\end{equation*}
To obtain $\om$, the first equation can be solved with the characteristic method \cite{diff-eq}, then $\a e^{\Om}$ can be obtained trivially. So, finally, one is able to write, also with the help of (\ref{eq-ee-rr-zz}), all the functions of the axisymmetric ansatz (\ref{axis-metric}) which characterise the electromagnetically charged Kerr-(A)dS metric, that is the Kerr-Newman-(A)dS:
\bea \label{charged-funct}
        \a &=&  \sin \t \sqrt{\Delta_e \Delta_\t}  \\ 
        e^{\Om/2} &=& \frac{\Delta_e - a^2 \Delta_\t \sin^2 \t}{\sqrt{\Delta_e \Delta_\t} \varrho^2 \sin \t}  \\
        \om &=&  \frac{a \sin^2 \t \left[ \Delta_e - \Delta_\t (r^2 + a^2) \right]}{a^2 \Delta_\t \sin^2 \t - \Delta_e}  \\
        e^{2\n} &=& \varrho^2 \sqrt{\a}  \qquad.
\eea
The electromagnetic potential $A_\m$ can be derived by the definition of $\mathbf{\Phi}$ (\ref{Phi-def}) and $\tilde{A_3}$ (\ref{A3tilde}):
\beq \label{am}
         A_\m = \left( \ \frac{e r -g a \cos \t}{r^2 + a^2 \cos^2 \t} \ ,\ 0 \ , \ 0 \ , \ \frac{-era\sin^2\t + g \cos \t (r^2+a^2)}{r^2 + a^2 \cos^2 \t} \ \right) \quad .
\eeq
It is also worth writing explicitly the coordinate transformation between (r, z) and ($r,\t$):
$$ \text{r}= \int \frac{dr}{\sqrt{ r^2 + a^2 - \frac{\Lambda}{3} r^4 - \frac{\Lambda}{3} r^2 a^2 - 2Mr + e^2 + g^2}}  \quad , \qquad \text{z}=\int \frac{d\t}{1 + \frac{\Lambda}{3} a^2 \cos^2\t}  $$
This pair of complex potential fields $\mathbf{\Phi}$ (\ref{Phi}) and $\Er$ (\ref{Er}), defined through (\ref{charged-funct})-(\ref{am}), satisfy the electrovacuum complex equations (\ref{ee-ernst-lambda})-(\ref{ee-ernst-lambda}) whenever $p^2+q^2+|Q|^2=1$ (with the usual uncharged $\xi$).\\
A similar procedure, even computationally easier because $q=0$, can be done in the static case to obtain the Reissner-Nordstrom-(A)dS metric from the Schwarzschild-(A)dS ``seed". \\

\section{New solution: Magnetic universe with cosmological constant}
\label{magn-universe}
 
In this section we want to perform a more significant test on the validity of the formalism developed in section \ref{complex-potencial}, applying it in the unknown ground of new solutions. \\
Is well known that the Melvin magnetic universe is a static, non-singular, cylindrical symmetric space-time in which there exist an axial magnetic field aligned with the z-axis. It represents a universe containing a parallel bundle of electromagnetic flux held together by its own gravitational field. This metric can be easily obtained by the Ernst technique when the cosmological constant is null \cite{ernst-magnetic}, just using the Minkowski space-time as a seed and applying a Harrison transformation. \\
Here, our aim is to generalise Melvin's solution in presence of the cosmological constant. In this case the axial Killing vector under consideration is $\p_\varphi$; hence the  relative axially symmetric static line element is:
\beq \label{axis-static-magn}
                          ds^2 = \a e^{-\Omega/2} dt^2 - \a e^{\Omega/2} d\varphi^2 + \frac{e^{2 \nu}}{\sqrt{-\alpha}} \left( d\text{r}^2 + d\text{z}^2 \right) \ \ ,
\eeq
where $\alpha$ is negative\footnote{To be as much close as possible to the original Ernst's notation of \cite{ernst-magnetic}.}. Then let us consider, as a ``seed", a uncharged metric without any curvature singularity, solution of Einstein equations with cosmological constant:
\beq
\label{mag-seed}
                    ds^2 = \left(1+\frac{\rho^2}{4}\right)^2 \left[ -dt^2 + d\text{z}^2 + \frac{\rho^2 \ d \rho^2 }{k\left(1+\frac{\rho^2}{4}\right) - \frac{4\Lambda}{3}\big(1+\frac{\rho^2}{4}\Big)^4} \right] + \frac{k\left(1+\frac{\rho^2}{4}\right) - \frac{4\Lambda}{3}\left(1+\frac{\rho^2}{4}\right)^4}{\left(1+\frac{\rho^2}{4}\right)^2} d\varphi^2 \ \ , 
\eeq
where $\text{r}=\int \alpha(\rho)^{-1} \rho d\rho$. 
For a certain value of the constant parameter $k=4\Lambda/3$, also angular singularities may be removed, so in that case, the metric is completely regular and thus it can be considered an AdS soliton. On the other hand when $k=0$ the Riemannian curvature becomes constant, so one remains with a locally (A)dS space-time. Similarly the asymptotic, for large $\rho$, results locally (anti) de Sitter. But for the moment we prefer to leave $k$ generic. \\
The complex potential associated with this choice of Killing vector $\p_\varphi$ is:
\beq								
      \Er_0 = f_0 = \a_0 e^{\Om_0/2} = -\frac{k}{1+\frac{\rho^2}{4}} + \frac{4\Lambda}{3} \left(1+\frac{\rho^2}{4} \right)^2       \quad .
\eeq
As in the previous black hole example, the charged function $f$ is obtained summing to $\Er_0$ the electromagnetic complex potential of the known $\Lambda=0$ case, that is the one of the Melvin or the Ernst solution, which in this coordinates is:
$$ \mathbf{\Phi} = \frac{\pm 2B}{1+\rho^2/4} \quad , $$
to get:
\beq
      f = \text{Re}(\Er + |\mathbf{\Phi}|^2) = -\frac{k}{1+\frac{\rho^2}{4}}  + \frac{4\Lambda}{3} \left(1+\frac{\rho^2}{4} \right)^2 + \frac{4 B^2}{\left(1+\frac{\rho^2}{4} \right)^2}     \quad .
\eeq
Next using (\ref{compl-em}) and (\ref{eq-ee-rr-zz}) is possible to find $\a , \Om$ and $\nu$:
\bea
      \alpha(\rho)     & = & \ -\sqrt{k\left(1+\frac{\rho^2}{4}\right)  - 4 B^2  - \frac{4\Lambda}{3}\left(1+\frac{\rho^2}{4}\right)^4} \nn  \\ 
      e^{\Om/2}(\rho)  & = & \ -\alpha(\rho) \left(1+\frac{\rho^2}{4}\right)^{-2}\\
      e^{2\nu}(\rho)   & = & \  \sqrt{-\alpha(\rho)} \left(1+\frac{\rho^2}{4}\right)^2 \quad .  \nn
\eea
Thus we have managed to charge the seed (\ref{mag-seed}), which reads:
\beq  \label{massive-magn-universe}
      ds^2 = \left(1+\frac{\rho^2}{4}\right)^2 \left[ -dt^2 + d\text{z}^2 + \frac{\rho^2 \ d \rho^2 }{\alpha^2(\rho)} \right] + \frac{\alpha^2(\rho)}{\left(1+\frac{\rho^2}{4}\right)^2} d\varphi^2 
\eeq
Which is supported by an electro-magnetic Maxwell potential $A_\m=\left[0,0,0,\pm\frac{2B}{1+\rho^2/4}\right]$.\\
The Melvin universe with cosmological constant can be obtained demanding regularity to the previous metric, as it hapens in the case without cosmological constant, thus costraining the $k$ parameter. If one expands (\ref{massive-magn-universe}) around $\rho \approx 0$, $k$ has to be fixed to $4B^2+4\Lambda/3$,  in order to avoid conical singularity. A closer aesthetically similarity with the usual ($\Lambda=0$) case is realised rescaling the coordinates such that $(\rho,\varphi) \rightsquigarrow (\bar{\rho} = \rho/B,\ \bar{\varphi}=\sqrt{B^4-B^2 \Lambda} \varphi)$:
\bea \label{lambda-melvin}
ds^2 &=& \left(1+\frac{B^2\bar{\rho}^2}{4}\right)^2 \left[ -dt^2 + d\text{z}^2 + \frac{d \bar{\rho}^2 }{1-\frac{\Lambda}{3} \left( \frac{3}{B^2} + \frac{3\bar{\rho}^2}{2} + \frac{B^2\bar{\rho}^4}{4} + \frac{B^4\bar{\rho}^6}{64} \right)}  \right] + \nn \\ 
     & & + \frac{1-\frac{\Lambda}{3} \left( \frac{3}{B^2} + \frac{3\bar{\rho}^2}{2} + \frac{B^2\bar{\rho}^4}{4} + \frac{B^4\bar{\rho}^6}{64} \right)}  {\left(1-\frac{\Lambda}{B^2}\right) \left(1+\frac{B^2\bar{\rho}^2}{4}\right)^2} \ \bar{\rho}^2 \ d\bar{\varphi}^2
\eea
This represents a generalisation of the Melvin magnetic universe in presence of negative cosmological constant\footnote{While for positive values of the cosmological constant the metrics's signature may change at large $\bar{\rho}$.
}. In fact, eventually, the standard Melvin solution is easily recovered setting $\Lambda=0$. Note that in this particular case is possible to switch off the magnetic field 
$  A_\m=\left[0,0,0,\pm \frac{\bar{\rho}^2 B^2/2}{\sqrt{B^2-\Lambda}(1+B^2\bar{\rho}^2/4)}\right] $
just in the limit of vanishing cosmological constant, obtaining the Minkowski space-time. Anyway a null magnetic field with non-null $\Lambda$ can be considered in the more general metric (\ref{massive-magn-universe}), trivially recovering the seed (\ref{mag-seed}). \\
Therefore, also in this case, the formalism elaborated in section \ref{complex-potencial} is useful to generate a charged solution starting from an uncharged seed, exactly in the same way as be done in the most relevant black hole case of section \ref{ads-bh}. Clearly, in order the method to be completely general, so to be able to work in any circumstance, a description of the whole symmetries of the Ernst equations still have to be carefully performed, as have been done in case of vanishing cosmological constant (for a review see \cite{ernst3}). This would stretch the technique developed in this paper to his maximum potentials.

\subsection{Rotating cosmological electro-magnetic universe}

For sake of generality let us present in this section a further generalisation of the metric (\ref{massive-magn-universe}). We are interested in adding rotation and electric charge. Summing the electric field, when the magnetic is already switched on, is somehow trivial because of the self-duality, in four dimension, between electric an magnetic field. Moreover (\ref{massive-magn-universe}) is more than just axisymmetric, it is cylindrical, which makes adding rotation on $\mathcal{S}^1$ a easy task. We end up with the stationary metric:
\bea \label{melvin-general}
     ds^2 &=& \left(1+\frac{\text{Q}^2 \rho^2}{4}\right)^2 \left[-(dt - \omega d \varphi)^2 + d\text{z}^2 + \frac{\rho^2 d \rho^2}{\frac{k}{\text{Q}^4}\left(1+\frac{\text{Q}^2\rho^2}{4}\right) - 4 \text{Q}^{-2} - \frac{4\Lambda}{3 \text{Q}^4}\left(1+\frac{\text{Q}^2\rho^2}{4}\right)^4} \right] + \nn \\
 & & \qquad \qquad \qquad \qquad  + \frac{\frac{k}{\text{Q}^4} \left(1+\frac{\text{Q}^2\rho^2}{4}\right) - 4 \text{Q}^{-2} - \frac{4\Lambda}{3 Q^4}\left(1+\frac{\text{Q}^2\rho^2}{4}\right)^4}{\left(1+\frac{\text{Q}^2\rho^2}{4}\right)^2} (d\varphi - \textrm{{\small $\upomega$}}  dt)^2  
\eea
whose electromagnetic potential is
$$ \qquad \qquad \qquad \qquad \     A_\m = \left[ \frac{\textrm{{\small $\upomega$}} \rho^2 B/2}{1+\frac{\text{Q}^2\rho^2}{4}} \ + E \ \text{z} ,\ 0 , \ 0 , \ -\frac{\rho^2 B/2}{1+\frac{\text{Q}^2\rho^2}{4}} - \ \omega \  E \ \text{z}  \right] \ \ .$$ 
The electric and magnetic constant parameters,  respectively $E$ and $B$, are related by the constraint $\text{Q}^2=E^2+B^2$. This is a stationary solution of Einstein-Maxwell-Lambda theory with angular momentum proportional to $\omega$. Again it is possible to avoid the conical singularity when $k=4(E^2+B^2)+4\Lambda/3 $, thus a rotating Melvin universe in presence of cosmological constant (which can consistently even be null) is retrieved. \\
It is interesting to observe that this metric (\ref{melvin-general}), when double analytically continued such that $t\rightarrow i\phi , \ \varphi \rightarrow i\tau$, represents a charged black brane, black string or toroidal black hole (depending on the angular identifications). The parameter $k$ is mapped into the mass parameter of the planar black hole, the electric and magnetic charges exchange one with the other,
In this picture the Melvin universe with cosmological constant (\ref{lambda-melvin}) corresponds to a double wick rotated electrically charged black string where the mass parameter is fine tuned to the electric charge.

\section{Comments and Conclusions}

In this paper we generalise Ernst's solution generating technique to be able to add electromagnetic charge to axisymmetric space-times in the theory of general relativity with cosmological constant. This represents both an extension of the method in \cite{ernst2}, because we are in presence of the cosmological constant, and at the same time is a generalisation of \cite{charmo1} because we add electromagnetic charge. This procedure can smoothly recover both \cite{ernst2} and \cite{charmo1} in the appropriate limit, that is $\L\rightarrow0$ and $Q\rightarrow0$, respectively. \\
As usual this technique is based on complex potential formalism and it takes advantages of integrability of the Einstein-Maxwell field equations to map the problem to a more tractable effective complex system. The Lagrangian of this effective system is presented and its symmetries are studied.\\
As a prototypical example we show how to add electromagnetic charge to the stationary rotating black hole of the theory considered: It is explicitly shown how to pass from the Kerr-(A)dS metric to the Kerr-Newman-(A)dS. \\
Moreover we take advantage of this formalism to extend to the cosmological constant case the Melvin solution which describes a static magnetic universe. Further generalisation such as the rotating and asymptotically (A)dS magnetic universe are find out.  \\
Of course the procedure is generic and can be applied, in principle, to any axisymmetric space-time, whenever set in the realm of General Relativity with cosmological constant (of any positivity). One has just to start from the uncharged metric one is interested in and treat it as a seed.\\
This formalism can be extended to other dimensions pretty directly, as done in the case without cosmological constant in \cite{ida} and \cite{iguchi}\footnote{Note that the extension of the technique does not guarantee find the hoped solution, in fact in \cite{ida} the charged rotating black hole is not found.}. While the generalisation to different theory of gravity such as the ones which comprise higher order in Riemann tensor seems more laborious. \\
The importance of a better understanding the solution generating formalism in presence of the cosmological constant are multiple. The most straightforward (but yet unexplored) application is probably to find a Schwarzschild-(A)dS black hole in a magnetic universe. Also interesting it may be discovering a cosmological constant extension of the Tomimatsu-Sato metrics. Between the most important applications there are maybe the search for (A)dS black ring; actually this was the main initial motivation of the author. Some attempts are made in four and five dimensions in \cite{asto-ring} and \cite{cald-empa-ring}, but a regular, asymptotically (A)dS, exact, ring solution is still unknown. To reach this scope a careful analysis of the inner symmetries of the effective complex action have to be carry out, works in this direction are in progress.

\section*{Acknowledgements}
\small
I would like to thank Fabrizio Canfora, Julio Oliva, Patricia Ritter, Ricardo Troncoso, Minas Tsoukalas and Jorge Zanelli for fruitful discussions; I am especially grateful to Eloy Ayon-Beato and Hideki Maeda for all their precious comments and suggestions. \\
\small This work has been funded by the Fondecyt grant 3120236. The Centro de Estudios Cient\'{\i}ficos (CECS) is funded by the Chilean Government through the Centers of Excellence Base Financing Program of Conicyt. CECS is also supported by a group of private companies which at present includes Antofagasta Minerals, Arauco, Empresas CMPC, Indura, Naviera Ultragas and Telef\'{o}nica del Sur.
\normalsize


\end{document}